\documentclass[preprint,nofootinbib,nobibnotes,groupaddress,preprintnumbers,aps,prd]{revtex4}
\usepackage{bm,epsfig,mathrsfs,amsmath,amssymb,graphicx}
\usepackage{epsfig,amsmath,graphicx,amssymb}
\usepackage{graphicx}
\usepackage{dcolumn}
\usepackage{bm}
\def\dbar{{\mathchar'26\mkern-12mu d}}

\begin{document}

\title{Efficiency at maximum power output of quantum heat engines under finite-time operation}

\author{Jianhui Wang}
\email{physwjh@gmail.com} \affiliation{ Department of Physics,
Nanchang University, Nanchang 330031, China}
\author{Jizhou He}
\affiliation{ Department of Physics, Nanchang University, Nanchang
330031, China}
\author{Zhaoqi Wu}
\affiliation{ Department of Mathematics, Nanchang University,
Nanchang 330031, China}

\begin{abstract}

We study the efficiency at maximum power, $\eta_m$, of irreversible
quantum Carnot engines (QCEs) that perform finite-time cycles
between a hot and a cold reservoir at temperatures $T_h$ and $T_c$,
respectively. For QCEs in the reversible limit (long cycle period,
zero dissipation), $\eta_m$ becomes  identical to  Carnot efficiency
$\eta_{_C}=1-\frac{T_c}{T_h}$.  For QCE cycles in which nonadiabatic
dissipation and time spent on two adiabats are included, the
efficiency $\eta_m$ at maximum power output is bounded from above by
$\frac{\eta_{_C}}{2-\eta_{_C}}$ and from below by
$\frac{\eta_{_C}}2$. In the case of symmetric dissipation, the
Curzon-Ahlborn efficiency $\eta_{_{CA}}=1-\sqrt{\frac{T_c}{T_h}}$ is
recovered under the condition that  the time allocation between the
adiabats and the contact time with the reservoir satisfy a certain
relation.

 %We study the corresponding
%quantum Carnot cycle of a
%simple two-state model with a one-dimensional power-law potential,
%and find that the expression of efficiency for the quantum Carnot
%cycle is shown to be identical to the classical one.
%We prove that under the same conditions, the efficiency is bounded
%from above the Carnot efficiency, even quantum dynamics is
%reversible.


PACS number(s): 05.70.Ln, 05.30.-d
\end{abstract}

\maketitle
\date{\today}
%%%%%%%%%%%%%%%%%%%%%%%%%%%%%%%%%%%%%%%%%%%%%%%%%

\section {introduction}
The concept of Carnot efficiency  is of
paramount importance in thermodynamics, since the Carnot cycle is
the most efficient heat engine cycle allowed by physical laws. When
the thermodynamic second  law states that not all the supplied heat
is applied to producing work, the Carnot efficiency presents the
limiting value on the fraction of the heat which can be so used.
Although the quasistatic Carnot cycle has the highest efficiency, it
outputs zero power because it takes infinite time to output a finite
amount of work. By contrast, Curzon and Ahlborn \cite{Cur75}
considered a finite-time Carnot cycle under the assumption of
endoreversibility that irreversible processes occur only through
these heat exchanges, they obtained the efficiency $\eta_{_{CA}}$ at
maximum power output as
\begin{equation}
\eta_{_{CA}}=1-\sqrt{\frac{T_c}{T_h}}, \label{etth}
\end{equation}
where $T_h$ and $T_c$ are the temperatures of the hot and cold heat
reservoirs, respectively. The Curzon-Ahlborn (CA) paper has
triggered the development of research into finite time
thermodynamics \cite{Ber00, Esp09, Esp10, Van05, He02, Che06, Cal85,
Gev92, Bej97, Kos84, Fel00, Chen89, Tu08, The08, Sch08,Gom06, Esp11,
Rez06}. The above $\eta_{_{CA}}$ is usually called the CA
efficiency, describing the efficiency of several engine models
\cite{Esp09, Sch08, Izu08, The08} and of actual thermal plants
\cite{Cur75, Bej97, Cal85, Esp09, Esp10, Gev92} very well. The CA
efficiency has been found to be a universal result in the case of
the low, asymmetric dissipation, by optimizing power output with
respect to time allocation when time durations in adiabats and
nonadiabatic phenomenon were ignored \cite{Esp10}.

Great efforts have been devoted to the study of quantum heat engines
\cite{Gev92, Scu11, Scu10, Scu03,  Ben00, He02, Che06, Wang07,
Quan06,  Oku11, W840, Kim11, Fel00,Rez06}, beginning with the
concept of quantum heat engine introduced by Scovil and
Schulz-DuBois \cite{Sco59}. Quantum heat engines differ from
classical counterparts mainly in the  following three respects: (i)
the working substance is composed of quantum matter such as spin
systems \cite{Wang07, Fel00, Gev92, He02, Che06}, harmonic
oscillator systems \cite{Che06, Gev92, Rez06}, two-level or
multilevel systems \cite{Quan06,  Oku11, Ben00, W840}, cavity
quantum electrodynamics systems \cite{Quan06, Scu03, Scu11, Scu10},
\emph{etc}. (ii) The state of the system is depicted by a
quantum-mechanical operator, and the thermodynamic observables are
associated with the expectation of values of operators \cite{Gev92,
Rez06, Fel00}. (iii) Quantum equations of motion are used to
describe the time evolution of the observables in quantum heat
engines, which can avoid  using phenomenological heat transfer laws
\cite{Gev92, Rez06, Fel00}.
%Quantum heat engines are
%usually characterized by three attributes: the working substance,
%the cycle of operation, and the dynamics that govern the cycle.
%The
%working substance of a quantum heat engine includes various quantum
%systems such as spin systems, harmonic oscillator systems
%\cite{He02,Che06, Feld00, Wu06, Wang07}, two-level or multilevel
%systems \cite{Quan07, Ben00}, cavity quantum electrodynamics systems
%\cite{Quan06, Scu03}, coupled two-level systems \cite{Hen07}, Finite
%time thermodynamics allows us to optimize power as well as
%efficiency.

  The previous literature discussed  the heat engine models in the
sudden limit in which the adiabatic process is a spontaneous
switching and thus the time allocation on adiabats is negligible
\cite{Esp10, Sch08, Tu08}. Thus it is significant to study more
general models in which the ``adiabatic"  process ( we take the two
corresponding processes as two quantum ``adiabats" throughout the
paper) takes finite time as well become nonadiabatic \cite{Nak11}.
During a quantum adiabatic process, the variation of the
eigenspectrum (quantum state) of the system must be  so slow that
the quantum adiabatic theorem \cite{Foc28, Oku11, W840, Quan06} can
apply. Otherwise, nonadiabatic dissipation (e.g., inner friction
\cite{Fel00, Wang07, Rez06}) occurs because of rapid change in the
energy level structure of the quantum system. Particularly
nonadiabatic dissipation has been found to have a profound influence
on the performance of quantum heat engines \cite{Fel00, Wang07,
Rez06}. Including nonadiabatic dissipation is therefore essential
for more realistic models of quantum heat engines.

In this paper,   we study the efficiency at maximum power output of
QCEs performing finite time cycles, in which the time of any adiabat
and nonadiabatic dissipation are considered.  We assume that the
external parameter affecting the energy spectrum varies at a small
but fixed speed which, however, may not be slow enough and thus to
cause nonadiabatic phenomenon.  We derive the cycle period that
consists of times spent both on the two quantum isotherms and on the
two quantum adiabats. We show that the efficiency at maximum power
output converges to an upper and  a lower bound in the limits of
extremely asymmetric dissipation.  Based on the  low-dissipation
assumption that the irreversible entropy production in a
thermodynamic process is inversely proportional to the time required
to complete that process, our approach similar to that of the
classical thermodynamics predicts that the CA efficiency turns out
to be an exact and universal property for QCEs operating under the
conditions that  the dissipation is symmetric and the    time
allocation between the adiabats and the contact time with the
reservoir satisfy   a certain relation.

%Physically, friction is the result of nonadiabatic phenomena which
%are the result of the rapid change in the energy level structure of
%the system.
%The rest of the paper is as follows. We capitulate the structure of
%a reversible two-level engine model  of a single atom in a harmonic
%trap in Sec. \ref{model}. In Sec. \ref{opt}, we study the
%optimization on power output, and we analyze the optimal ranges of
%the efficiency and of the engine structure. Finally, we present our
%conclusions in Sec. \ref{con} .

\section {Efficiency at maximum power output}

We consider a quantum system whose Schr\"{o}dinger's equation is
given by ${H}|u_n\rangle=E_n|u_n\rangle$, where ${H}$, $|u_n\rangle$
and $E_n$ are the Hamiltonian of the system, its $n$th eigenstate
and eigenenergy, respectively. The internal energy $U$ reads
$U=\sum_nE_nP_n$, where $P_n$ is the mean occupation probability of
the $n$th eigenstate and obeys the canonical distribution
$P_n=\frac{1}Ze^{-E_n/k_BT}$ in equilibrium, with the canonical
partition function $Z=\sum_ne^{-E_n/k_BT}$. Derivation of $U$ leads
to the first quantum thermodynamic law
$dU=\sum_{n}E_ndP_n+\sum_{n}P_ndE_n$. Analogous to the classical
thermodynamic first law, the first law of thermodynamics in
quantum-mechanical systems is \cite{Quan06, Kim11, W840}  $dU=\dbar
Q+\dbar W$, in which  $\dbar Q=\sum_{n}E_ndP_n$ and $\dbar
W=\sum_{n}P_ndE_n$ depict the heat exchange and work done,
respectively, during a thermodynamic process. Note that
$\sum_{n}E_ndP_n$ is associated with the heat exchange because
$\dbar Q=TdS$ with the entropy $S=-k_BP_n\ln P_n$. Motivated by the
definition of the generalized force $F$ for a classical system, we
define analogously the force for a quantum system as $F=\sum_n P_n
\frac{\partial{E_n(X)}}{\partial X}$,  where $X$ is the external
parameter (generalized coordinate corresponding to the force $F$)
\cite{Quan06, note1, W840, Kim11}. Here the force $F$ and
generalized coordinate $X$ are state variables \cite{note1, Kim11}
and quantum versions of the classical pressure $P_r$ and volume $V$,
respectively.
\begin{figure}[h]
\includegraphics[width=230pt]{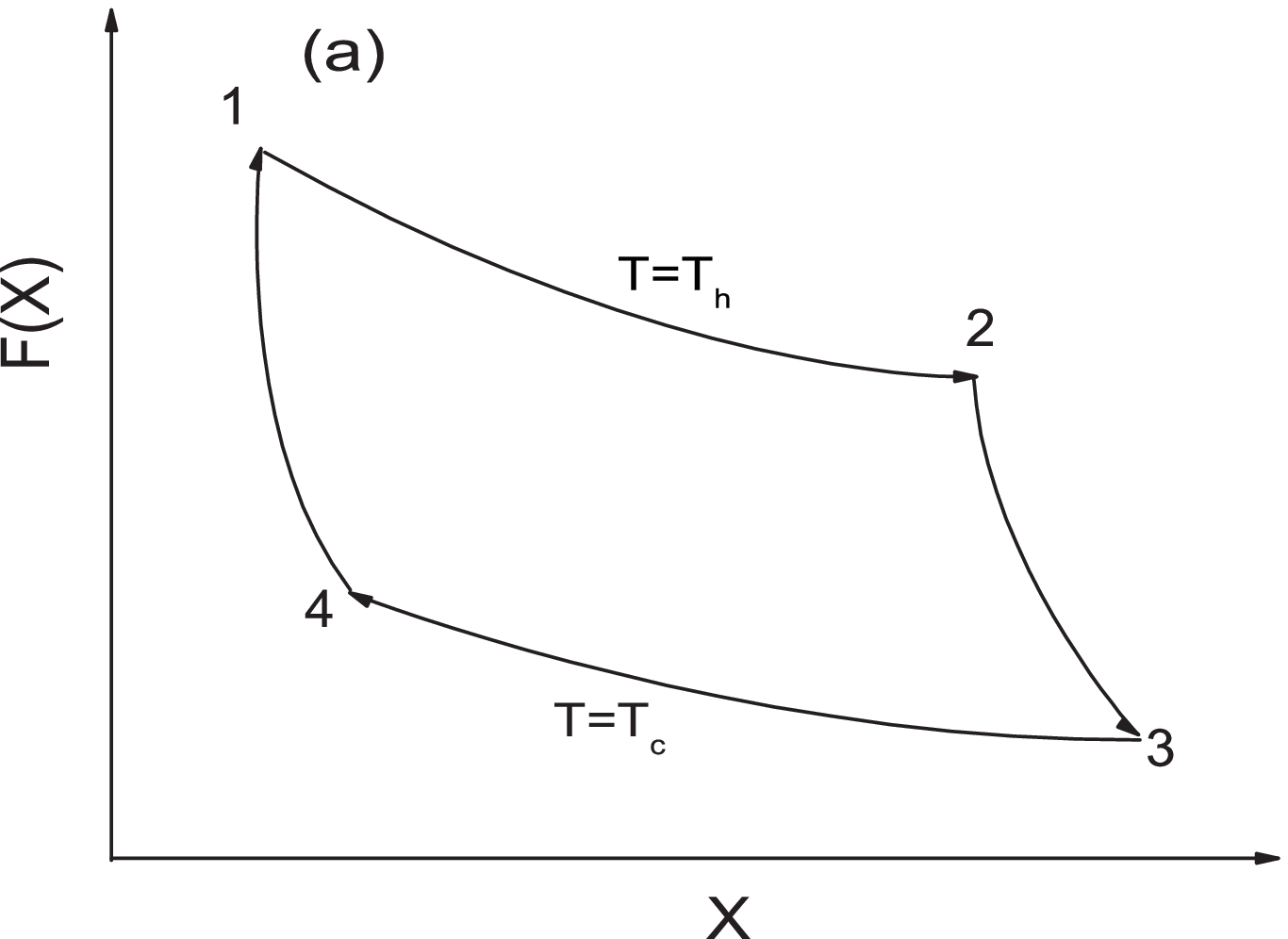}
\includegraphics[width=230pt]{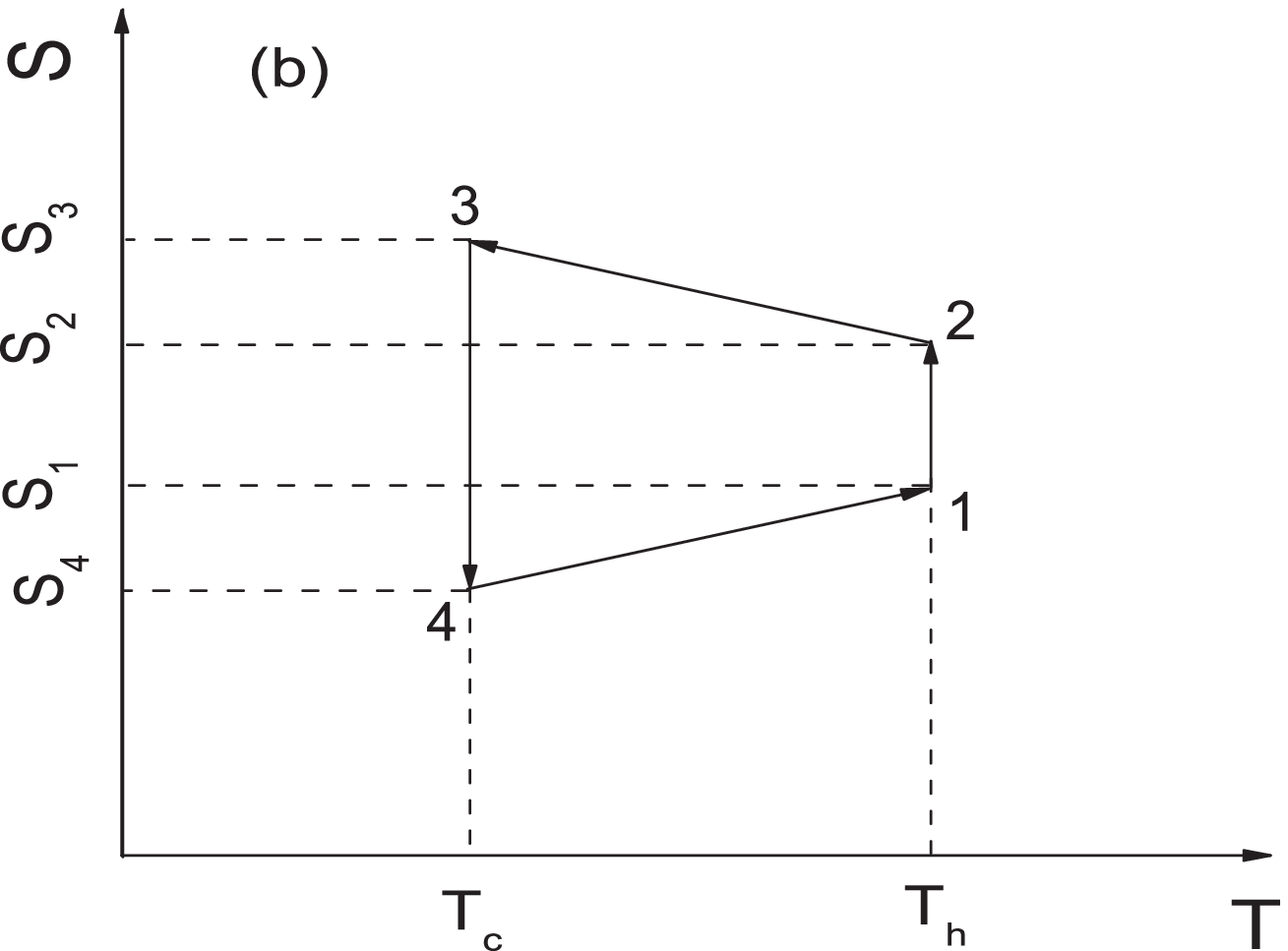}
\caption{ Schematic diagram of an irreversible QCE cycle in the
plane of the external parameter $X$ and force $F(X)$ (a) and of the
Temperature $T$ and entropy $S$ (b). The values of the external
parameter $X$ and of the entropy $S$ at the four special instants
are indicated.} \label{fll}
\end{figure}

The generalized force $F(X)$ alters the generalized coordinate $X$
that affects the eigenspectrum of the system during a thermodynamic
process. The quantum Carnot cycle
$1\rightarrow2\rightarrow3\rightarrow4\rightarrow1$ is drawn in the
$(F, X)$ plane, which is the quantum version of the classical $(P_r,
V )$ plane (See. Fig. \ref{fll}a).  During two quantum isothermal
processes $1\rightarrow2$ and $3\rightarrow4$, the working substance
is coupled to a hot and a cold heat reservoir at constant
temperatures $T_h$ and $T_c$, respectively.  We apply $\dbar Q=TdS$
directly to the calculation of the heat exchange $\dbar Q$ in any
quantum isothermal process. Let $S(X_i)$ and $X_i$ be the entropies
and the external parameters at the instants $i$ with $i=1,2,3, 4$,
the heat amount $Q_h$ absorbed from the hot reservoir and the heat
amount $Q_c$ released to the cold reservoir are, respectively,
$Q_h=T_h[S(X_2)-S(X_1)]$ and $Q_c=T_c |[S(X_4)-S(X_3)]|$. When the
Carnot cycle is reversible, the system couples to the heat reservoir
for a sufficiently long time until the system remains
infinitesimally close to equilibrium all along the cycle, and in
order for the adiabatic theory to remain valid, the time scale of
the change of the quantum state must be much larger than that of the
dynamical one, $\sim E/{\hbar}$ \cite{Foc28, Oku11, W840}. For the
reversible cycle where $S(X_2)=S(X_3)$ and $S(X_1)=S(X_4)$, we
recover the Carnot efficiency $\eta_{_C}=1-\frac{T_c}{T_h}$, which
is independent of the properties of the quantum working substance.
We denote by $t_{12}$ ($t_{34}$)  the time durations during which
the system is coupled to the hot (cold) reservoir along a cycle. In
the branch $2\rightarrow3$ ($4\rightarrow1$), the working substance
is decoupled from the hot (cold) reservoir, and the entropy changes
from $S_2$ to $S_3$ ($S_4$ to $S_1$) during a period $t_{23}$ (
$t_{41}$), as shown in Fig. \ref{fll}b, where $S_i\equiv S(X_i)$
with $i=1, 2, 3, 4$.

Let us consider a QCE under finite-time operation.
 Finite-time  cycles  move
the system away from the equilibrium, leading to irreversibility of
the engine. Although the system needs no close to equilibrium during
the isothermal process,  the system remains in an equilibrium state
with the heat reservoir at special instants $i$ with $i=1, 2, 3, 4$.
Under such a circumstance,  the thermodynamic quantities of the
system$-$in particular the entropy$-$are well defined at these
instants.  During the processes $2\rightarrow3$ and $4\rightarrow1$,
the rapid change (compared with time scale, $\sim E/\hbar$) in the
energy level structure of the system results in quantum nonadiabatic
phenomenon.  We emphasize that in any quantum thermodynamic process
in which the low dissipation exists the system relaxation is assumed
to be fast compared to the time of the process \cite{Esp10}. During
a quantum isothermal (adiabatic) process, the entropy production
caused by weak dissipation can be written  as ${\Sigma_h}/{t_{12}}$
or ${\Sigma_c}/{t_{34}}$ (${\Sigma_a}/{t_{23}}$ or
${\Sigma_a}/{t_{41}}$), since the reversible regime is approached in
the infinite time durations. Thus, the real heat exchanges $Q_h$ and
$Q_c$ are
 $Q_h=T_h\Delta S-T_h\frac{\Sigma_h}{t_{12}}$ and $Q_c=T_c\Delta
S+T_c\frac{{\Sigma_c}}{t_{34}}+T_c(\frac{{\Sigma_a}}{t_{23}}+\frac{{\Sigma_a}}{t_{41}})$,
with $\Delta
S=S(X_2)-S(X_1)=|[S(X_4)-S(X_3)]|-(\frac{{\Sigma_a}}{t_{23}}+\frac{{\Sigma_a}}{t_{41}})$.
According to the first law of the thermodynamics, the work $W$ per
cycle is determined by
\begin{equation}
W=Q_h-Q_c={(T_h-T_c)}\Delta
S-T_h\frac{\Sigma_h}{t_{12}}-T_c\frac{\Sigma_c}{t_{34}}-T_c(\frac{{\Sigma_a}}{t_{23}}+\frac{{\Sigma_a}}{t_{41}}).
\label{wqs1}
\end{equation}
Let $v(t)$ and $\tau$ be the speed of the change of $X$ and the
cycle period, respectively. Then we have
\begin{eqnarray}
X_0&=&(X_2-X_1)+(X_3-X_2)+(X_3-X_4)+(X_4-X_1)\nonumber\\
 &=& 2(X_3-X_1)=\int_0^\tau v(t)dt=\bar{v}\tau, \label{x0au}
\end{eqnarray}
where $\bar{v}$ is the average speed of the change of $X$. The
displacement of $X$ after a single cycle is zero and  thus $X$ is a
state variable, though the total change per cycle $X_0$ is not equal
to zero. The times spent on two isothermal processes can be
expressed as $t_{12}=(X_2-X_1)/\bar{v}$ and
$t_{34}=(X_3-X_4)/\bar{v}$, respectively, while the times of two
adiabats are
$t_{23}=(X_3-X_2)/\bar{v}$ and $t_{41}=(X_4-X_1)/\bar{v}$.  %For instance, there
%are several special cases: (i) $\sigma=\alpha= 2$ for a box and a
%harmonic potential, $B=\frac{\pi^2\hbar^2}{2m}$ for a box potential,
%while $B=\frac{\hbar^2}{m}$ for a harmonic potential \cite{Xio02},
%with $m$ being the particle mass. (ii) $\sigma=\alpha= 1$ and
%$B={\pi \hbar c}$ for extremely relative particles in a  box
%potential \cite{Pat03}. (iii) $\sigma=\frac{4}3$, $\alpha=2$, and
%$B=\frac{\hbar^2}{m}$ for a quartic potential \cite{wepjd}.
Therefore, the power output $P=W/\tau$ and the efficiency
$\eta=W/Q_h$ are
\begin{equation}
P=\frac{\bar{v}}{2(X_3-X_1)}{[(T_h-T_c)}\Delta
S-\frac{\bar{v}T_h{\Sigma_h}}{(X_2-X_1)}-\frac{\bar{v}T_c{\Sigma_c}}{(X_3-X_4)}-
\frac{\bar{v}T_c{\Sigma_a}}{(X_3-X_2)}-\frac{\bar{v}T_c{\Sigma_a}}{(X_4-X_1)}],
\label{pf2r}
\end{equation}
and
\begin{equation}
\eta=\frac{{(T_h-T_c)}\Delta
S-\frac{\bar{v}T_h{\Sigma_h}}{(X_2-X_1)}-\frac{\bar{v}T_c{\Sigma_c}}{(X_3-X_4)}
-\frac{\bar{v}T_c{\Sigma_a}}{(X_3-X_2)}-\frac{\bar{v}T_c{\Sigma_a}}{(X_4-X_1)}}{{T_h}\Delta
S-\frac{\bar{v}T_h{\Sigma_h}}{(X_2-X_1)}}, \label{et21}
\end{equation}
respectively. Here $\Delta S$  is a state variable determined only
by the initial and final states of the  isothermal  process. The
generalized coordinates $X_i$ with $i=1,2,3,4$ , corresponding to
the system volume $V$ in the classical thermodynamics, are state
variables and independent of the detailed protocols. To specify the
time allocation at maximum power output, the values of $X_i$ as well
as the average speed $\bar{v}$ should be optimized. We will the
optimize power output $P$ over the average speed $\bar{v}$ and the
variables $X_i$ to obtain the time allocation during a cycle and
thus to determine the corresponding efficiency. We will assume, for
simplicity, that the initial value of the external parameter is a
constant, i.e, $X_1=X_1^0$. The maximum power is therefore found by
setting the derivatives of $P$ with respect to the average speed
$\bar{v}$ and $X_i$ with $i=2,3,4$ equal to zero.

The maximization conditions
$\frac{\partial{P}}{\partial{X_i}}|_{{X_i}=X_i^m}=0$ and
$\frac{\partial{P}}{\partial\bar{v}}|_{\bar{v}=\bar{v}_m}=0$ give
the physical solution.  The value of $X_3^m$ is determined by the
following equation
\begin{eqnarray}
(T_h-T_c)\Delta S &=&\bar{v}_m
T_c\Sigma_a(\frac{1}{X_3^m-X_2^m}+\frac{1}{X_4^m-X_1^m})
+\frac{\bar{v}_m T_c\Sigma_c}
 {X_3^m-X_4^m}+\bar{v}_m
T_c(X_3^m-X_1^0) \nonumber\\
&\times &[\frac{\Sigma_a}{(X_3^m-X_2^m)^2}
+\frac{\Sigma_c}{(X_3^m-X_4^m)^2}]+\frac{\bar{v}_mT_h\Sigma_h}{X_2^m-X_1^0},\label{thx1}
\end{eqnarray}
where
\begin{equation}
\bar{v}_m=\frac{1}2\frac{(T_h-T_c )\Delta S}{
(\frac{1}{X_3^m-X_2^m}+\frac{1}{X_4^m-X_1^0}){T_c\Sigma_a}
+\frac{1}{X_3^m-X_4^m}T_c\Sigma_c+\frac{1}{X_2^m-X_1^0}T_h\Sigma_h},
\label{baah}
\end{equation}
\begin{equation}
X_2^m=\frac{|T_h\Sigma_h X_3^m-T_c\Sigma_a
X_1^0|+(X_3^m-X_1^0)\sqrt{T_cT_h\Sigma_a\Sigma_h}}{|T_h\Sigma_h-T_c\Sigma_a|},
\label{x2maa}
\end{equation}
and
\begin{equation}
X_4^m=\frac{|\Sigma_a X_3^m-\Sigma_c X_1^0|+(X_3^m-X_1^0)\sqrt{
\Sigma_a\Sigma_c}}{|\Sigma_a-\Sigma_c|}. \label{x4maa}
\end{equation}
From Eqs. (\ref{thx1}) and (\ref{baah}), we find that the optimal
value of $X_3^m$  with fixed value of  $X_1^0$ is independent of the
value of the state variable $\Delta S$, as expected.  Substitution
of Eqs. (\ref{baah}), (\ref{x2maa}) and (\ref{x4maa}) into Eq.
(\ref{thx1}) leads to the fundamental optimal relationship between
$X_3^m$  and $X_1^0$ at maximum power output. Under  the assumption
that the value of $X_1$ is fixed at the start of the engine cycle,
Eq. (\ref{thx1}) can be done numerically for given values of entropy
production $\Sigma_a$, $\Sigma_c$ and $\Sigma_h$ along the specific
processes and of temperatures $T_h$ and $T_c$.  Once we have
obtained the optimal relationship between $X_1^0$
 and $X_3^m$
at maximum power output $P$, we can then determine the optimal
values of $X_i^m$ with $i=2, 4$,   and the average speed $\bar{v}_m$
by Eqs. (\ref{baah}), (\ref{x2maa}), and (\ref{x4maa}).

\begin{figure}[h]
\includegraphics[width=300pt]{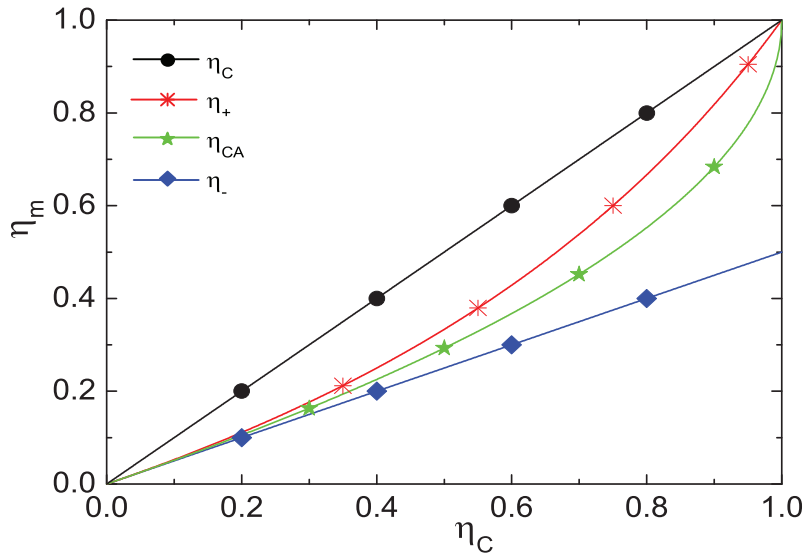}
\caption{ (Color online). Efficiency $\eta_m$ at maximum power as a
function of the Carnot value $\eta_{_C}$. The upper and lower bounds
of the efficiency, $\eta_+$ and $\eta_-$ given in Eq.
(\ref{bounds}), are represented by a red and a blue solid line with
starts  and squares, respectively. The CA efficiency $\eta_{_{CA}}$
is denoted by a green  solid line with 5-pointed stars, while the
Carnot efficiency $\eta_{_C}$ by a black solid line with circles.}
\label{etam}
\end{figure}

 Substituting Eq. (\ref{baah}) into Eq. (\ref{et21}),  we find
the expression for the efficiency at maximum power as follows:
\begin{equation}
\eta_m=\frac{1-\frac{T_c}{T_h}}{2-\frac{(T_h-T_c)}{(X_2^m-X_1^0)[(\frac{T_c}{X_3^m-X_2^m}+\frac{T_c}{X_4^m-X_1^0})
\frac{\Sigma_a}{\Sigma_h}+\frac{T_c}{X_3^m-X_4^m}\frac{\Sigma_c}{\Sigma_h}+\frac{T_h}{X_2^m-X_1^0}]}
}%\frac{(\frac{1}{X_3-X_2}+\frac{1}{X_4-X_1}
. \label{etha}
\end{equation}
Eq. (\ref{etha}) together with Eqs. (\ref{thx1}), (\ref{baah}),
(\ref{x2maa}), and (\ref{x4maa}), as one of our main results,
conveys the following physical features:

(i) The nonadiabatic dissipation is neglected, i.e.,
${\Sigma_a}\rightarrow 0$. In such a case, the limits
$\frac{\Sigma_c} {\Sigma_h}\rightarrow 0$ and $\frac{\Sigma_c}
{\Sigma_h}\rightarrow \infty$, lead to the result that the
efficiency $\eta_m$ at the maximum power approaches to the upper
bound $\eta_+\equiv\frac{\eta_{_C}}{2-\eta_{_C}}$ and to the lower
bound $\eta_-\equiv\frac{\eta_{_C}}2$, respectively. That is, the
efficiency $\eta_m$ at the maximum power satisfies the following
condition:
\begin{equation}
\frac{\eta_{_C}}2\equiv\eta_-\leq\eta_m\leq\eta_+\equiv\frac{\eta_{_C}}{2-\eta_{_C}}.
\label{bounds}
\end{equation}
In Fig. \ref{etam} we plot the efficiency (\ref{etha}) as a function
of $\eta_{_C}$ comparing $\eta_{_{CA}}$  with the upper and lower
bounds (\ref{bounds}). The lower and upper bounds, which are reached
in the completely asymmetric limits $\frac{\Sigma_c}
{\Sigma_h}\rightarrow 0$ and $\frac{\Sigma_c} {\Sigma_h}\rightarrow
\infty$,  are identical to the corresponding those derived in
different approaches \cite{Esp10}. However, unlike the previous
literature in which the times of two adiabats are ignored, the times
spent on the two adiabats in the quantum Carnot cycle are taken into
account. If the symmetric dissipation ${\Sigma_c} ={\Sigma_h}$ and
$\frac{X_2^m-X_1^0}{X_3^m-X_4^m}=\sqrt{\frac{T_h}{T_c}}$, i.e., the
time allocation to the hot and cold processes at maximum power:
\begin{equation}
\frac{t_{12}}{t_{34}}=\sqrt{\frac{T_h}{T_c}}, \label {frtc}
\end{equation}
we can recover the CA efficiency
$\eta_m=\eta_{_{CA}}=1-\sqrt{\frac{T_c}{T_h}}$  by using Eq.
(\ref{etha}). Result of consideration of symmetric dissipation
agrees with that obtained by optimizing power output with respect to
the times of the two isothermal processes \cite{Esp10, note2}.
%It is worth noting that consideration of symmetric dissipation
%reveals an important distinction between the classical and quantum
%Carnot cycles.  In the symmetric dissipation
%${\Sigma_c}={\Sigma_h}$, the condition of maximum power output for a
%classical Carnot cycle leads to Curzon-Ahlborn efficiency and
%$\frac{t_{12}}{t_{34}}=\sqrt{\frac{T_h}{T_c}}$. However, for a
%quantum Carnot cycle, the Curzon-Ahlborn efficiency is recovered
%under conditions that the weak dissipation is symmetric and  the
%time ratio of contact in heat baths is satisfied with Eq.
%(\ref{frtc}).

(ii) There is nonadiabatic dissipation, while   the dissipation in
at least one quantum isothermal process is not considered, namely,
${\Sigma_a}\neq 0$, and ${\Sigma_h}\rightarrow 0$ (or
${\Sigma_c}\rightarrow 0$). From Eq. (\ref{etha}) we find in this
case
\begin{equation}
\eta_m=\eta_{-}=\frac{\eta_{_C}}2, \label{etc2}
\end{equation}
 which is independent of
the  values  of both $\Sigma_a$ and $\Sigma_c$ (or $\Sigma_h$). The
lower bound, which is found in the case when  nonadiabatic
dissipation exists but
 dissipation vanishes in at least one isotherm, is particularly
interesting. Although it coincides with a reported universal lower
bound in Refs. \cite{Esp10, Tu08},  it is derived in the generalized
engine model with nonadiabatic phenomenon.   Physically, even in the
case when the two isothermal processes are reversible,  inclusion of
an arbitrary low dissipation in the adiabats reduces the efficiency
at maximum power output to half the Carnot value
$\frac{\eta_{_C}}2$.

%It also agrees with the upper bound obtained
%by optimizing with respect to the temperature of the hot reservoir
%[24]. Finally, it also arises in a model for the Feynman ratchet
%[26] [cf. Eq. (25)].

(iii) Dissipations in four quantum thermodynamic processes are
equal, i.e., $\Sigma_c/\Sigma_h=1$ and $\Sigma_a/\Sigma_h=1$. Let
$R_x\equiv(X_2^m-X_1^0)(\frac{1}{X_3^m-X_2^m}+\frac{1}{X_4^m-X_1^0}+\frac{1}{X_3^m-X_4^m})$,
in the limits $R_x\rightarrow 0$ and $R_x\rightarrow \infty$, the
efficiency $\eta_m$ corresponding to maximum power output converges
to  the upper bound $\eta_+=\frac{\eta_{_C}}{2-\eta_{_C}}$ and to
the lower bound $\eta_-=\frac{\eta_{_C}}2$, respectively. Here the
lower and upper bounds are equal to the corresponding those in
previous studies, but extended to the irreversible QCEs  in which
the time spent on two adiabats and nonadiabatic dissipation are
considered. According to Eq. (\ref{etha}), the CA efficiency
$\eta_{_{CA}}=1-\sqrt{\frac{T_c}{T_h}}$
 is achieved when the times spent on the four
quantum thermodynamic processes are distributed in such a way that
\begin{equation}
\frac{t_{12}{(\tau-t_{12})}}{t_{23}t_{34}t_{41}}=\sqrt{\frac{T_h}{T_c}},
\label{frtc2}
\end{equation}
where $t_{12}=(X_2^m-X_1^0)/\bar{v}_m,
t_{23}=(X_3^m-X_2^m)/\bar{v}_m, t_{34}=(X_3^m-X_4^m)/\bar{v}_m,$ and
$t_{41}=(X_4^m-X_1^0)/\bar{v}_m$.

\section{Conclusions}

In conclusion, we have determined efficiency at maximum power for a
QCE engine performing finite time cycles. To correctly describe the
irreversible QCEs the times spent on two adiabats and nonadiabatic
phenomenon have been taken into account. In the limits of extremely
asymmetric dissipation
 ($\frac{\Sigma_c}{\Sigma_h}\rightarrow 0$ and
$\frac{\Sigma_c}{\Sigma_h}\rightarrow \infty$, with
$\frac{\Sigma_a}{\Sigma_h}\rightarrow 0$), the efficiency at maximum
power output converges to an upper and a lower bound, coinciding
with the result obtained previously in different approaches. When
 dissipation in any isothermal process vanishes,
 the efficiency at maximum power output is equal to the lower
bound $\frac{\eta_{_C}}2$. For the QCE with the symmetric
dissipation but without nonadiabatic dissipation
($\frac{\Sigma_c}{\Sigma_h}\rightarrow 1$, while
$\frac{\Sigma_a}{\Sigma_h}\rightarrow 0$), we have derived CA
efficiency at maximum power output, only provided that the ratio of
the times of contact with two heat reservoirs satisfies the relation
given as in Eq. (\ref{frtc}). In the case of
$\frac{\Sigma_c}{\Sigma_h}\rightarrow 1$ and
$\frac{\Sigma_a}{\Sigma_h}\rightarrow 1$, we have also recovered CA
efficiency at maximum power output, if the time allocations of four
processes fulfill the condition in Eq. (\ref{frtc2}).

\emph{Acknowledgements:} We gratefully acknowledge support for this
work by the National Natural Science Foundation of China under Grant
Nos. 11147200 and 11065008. J. H. Wang also gratefully acknowledges
Z. C. Tu for his kind communications.


\begin{thebibliography}{99}
\bibitem{Cur75} F. Curzon and B. Ahlborn, Am. J. Phys. \textbf{43}, 22 (1975).
\bibitem{Esp09}
M. Esposito, K. Lindenberg, and C. Van den Broeck, Europhys. Lett.
\textbf{85}, 60010 (2009); B. Rutten, M. Esposito, and B. Cleuren,
Phys. Rev. B \textbf{80}, 235122 (2009); M. Esposito, R. Kawai, K.
Lindenberg, and C. Van den Broeck, Phys. Rev. E \textbf{81}, 041106
(2010).
\bibitem{Esp11}M. Esposito, R. Kawai, K. Lindenberg and C. Van den Broeck,  Europhys. Lett. \textbf{89}, 20003
(2010); N. Kumar, C. Van den Broeck, M. Esposito, and K. Lindenberg,
Phys. Rev. E \textbf{84}, 051134 (2011).
\bibitem{He02} J. Z.  He,  J. C. Chen, and  B. Hua,  Phys. Rev. E \textbf{65}, 036145 (2002).
\bibitem{Che06} F.  Wu,  L. G. Chen,  F. R. Sun,  C. Wu, and  Q. Li,
Phys. Rev. E \textbf{73}, 016103  (2006); F. Wu, L. G. Chen,  F. R.
Sun, C. Wu, and  F. Z. Guo, J. Appl. Phys. \textbf{99},
 054904 (2006); F. Wu, L. G. Chen, S. Wu, F. R. Sun, and  C. Wu, J. Chem.
Phys. \textbf{124}, 214702 (2006).
\bibitem{Cal85} H. B. Callen, \emph{Thermodynamics and an Introduction
Thermostatistics} (Wiley, New York, 1985), 2nd ed.

\bibitem{Bej97} A. Bejan,\emph{ Advanced Engineering Thermodynamics} (Wiley, New York,
1997), p. 377.

\bibitem{Esp10} M. Esposito, R. Kawai, K.  Lindenberg, and C. Van den Broeck, Phys.
Rev. Lett. \textbf{105}, 150603 (2010), and references therin.

\bibitem{Kos84} R. Kosloff, J. Chem. Phys. \textbf{80}, 1625 (1984).
\bibitem{Gev92}  E. Geva and R. Kosloff, J. Chem. Phys. \textbf{96}, 3054
(1992); E. Geva and R. Kosloff, J. Chem. Phys. \textbf{97},
4396(1992); E. Geva and R. Kosloff, Phys. Rev. E \textbf{49}, 3903
(1994); E. Geva and R. Kosloff, J. Chem. Phys. \textbf{102}, 8541
(1995).
\bibitem {Fel00}   T. Feldmann and  R. Kosloff
Phys. Rev. E \textbf{61}, 4774 (2000).
\bibitem{Rez06} Y. Rezek and
R. Kosloff, New J. Phys. \textbf{8}, 83 (2006).
\bibitem{Chen89} L. Chen and Z. Yan, J. Chem. Phys. 90, 3740 (1989)
\bibitem{Van05} C. Van den Broeck, Phys. Rev. Lett. \textbf{95}, 190602
(2005).

\bibitem{Ber00} R. S. Berry, V.A.
Kazakov, S. Sieniutycz, Z. Szwast, and A. M. Tsvilin, Thermodynamic
Optimization of Finite- Time Processes (John Wiley $\&$ Sons,
Chichester, 2000).
\bibitem{Sch08}  T. Schmiedl and U. Seifert,
Europhys. Lett. \textbf{81}, 20003 (2008).
\bibitem{Tu08}Z. C. Tu, J. Phys. A: Math. Theor. 41 (2008);
Y. Wang and Z. C. Tu, Phys. Rev. E. \textbf{85}, 011127 (2012); Y.
Wang and Z. C. Tu, arXiv:1110.6493v2.
\bibitem{The08} H. Then and A.
Engel, Phys. Rev. E \textbf{77}, 041105 (2008).
\bibitem{Gom06} A.
Gomez-Marin and J. M. Sancho, Phys. Rev. E 74, 062102 (2006).
\bibitem{Izu08} Y. Izumida
and K. Okuda, Europhys. Lett. \textbf{83}, 60003 (2008); Phys. Rev.
E \textbf{80}, 021121 (2009); Prog. Theor. Phys. Suppl.
\textbf{178}, 163 (2009).

\bibitem{Ben00}  C. M. Bender,  D. C. Brody, and B. K. Meister, J.
Phys. A: Math. Gen. \textbf{33}, 4427 (2000).

%\bibitem{Quan09}
%H. T. Quan,Y. X.  Liu, C. P. Sun, and F. Nori, Phys. Rev. E
%\textbf{76}, 031105 (2007).
%\bibitem{Scu03} M. O. Scully,  M. S. Zubairy, G. S. Agarwal, and H. Walther, Science
%\textbf{299}, 862  (2003).
%%\bibitem{Scu03}  M. O. Scully, S.
%%Zubairy, G. Agarwal, and H. Walther, Science \textbf{299}, 862
%%(2003).
%;  T. Feldmann and  R. Kosloff,
%Phys. Rev. E \textbf{68}, 016101 (2003);  T. Feldmann and R.
%Kosloff, Phys. Rev. E \textbf{70}, 046110 (2004).

%J. H. Wang,  J. Z. He, and Y. Xin, Phys. Scr. \textbf{75}, 227
%(2007).



%
%\bibitem{Abe11} .
%
\bibitem{Oku11} S. Abe and S. Okuyama, Phys. Rev. E \textbf{83}, 021121 (2011);
S. Abe, Phys. Rev. E \textbf{83}, 041117 (2011).
%
%
%\bibitem{Hen07}  M. J. Henrich,  G. Mahler, and  M.
%Michel,  Phys. Rev. E \textbf{75}, 051118 (2007).
%
%
\bibitem{W840}J. H. Wang, J. Z. He, and X. He, Phys. Rev. E  \textbf{84}, 041127 (2011);
J. H. Wang and J. Z. He, J. Appl. Phys. \textbf{111}, 043505 (2012).
%\bibitem{Per98} P. Perrot,\emph{ A to Z of Thermodynamics} (Oxford University Press,
%Oxford, 1998).
%\bibitem{Che98}J. Chen and Z. Yan, J. Appl. Phys. \textbf{84}, 1791 (1998);
%J. Chen and Z. Yan, J. Appl. Phys. \textbf{69}, 6245 (1991).
%\bibitem{Scuopt} M. O. Scully and M. S. Zubairy, \emph{Quantum Optics} (Cambridge
%                 University Press, Cambridge, 1997).

\bibitem{Quan06} H. T.  Quan,  P. Zhang, and  C. P.  Sun,   Phys. Rev. E \textbf{73}, 036122 (2006);
H. T.  Quan, Y. X.  Liu,  C. P. Sun, and  F. Nori, Phys. Rev. E
\textbf{76}, 031105 (2007); H. T. Quan, Phys. Rev. E \textbf{79},
041129 (2009).
\bibitem{Kim11} S. W. Kim, T. Sagawa,  S. De Liberato,  and M. Ueda,
Phys. Rev. Lett. \textbf{106}, 070401 (2011).

\bibitem{Wang07} J. H. Wang,  J. Z. He, and Y. Xin, Phys. Scr. \textbf{75}, 227
(2007).
\bibitem{Scu10} M. O. Scully, Phys. Rev. Lett. \textbf{104}, 207701 (2010); K. E. Dorfman, M. B. Kim,  and A. A.
Svidzinsky, Phys. Rev. E   \textbf{84}, 053829 (2011).
\bibitem{Scu11} M. O. Scully, K. R. Chapin, K. E. Dorfman, M. B. Kim, and A. A.
Svidzinsky, PNAS \textbf{108}, 15097 (2011).

\bibitem{Scu03} M. O. Scully,  M. S. Zubairy, G. S. Agarwal, and H. Walther, Science
\textbf{299}, 862  (2003).
\bibitem{Sco59}H. E. D. Scovil and E. O. Schulz-DuBois, Phys. Rev. Lett. \textbf{2}, 262
(1959).
\bibitem{Nak11} K. Nakamura,  S. K. Avazbaev, Z. A. Sobirov, D. U. Matrasulov,
and T. Monnai, Phys. Rev. E \textbf{83}, 041133 (2011).

\bibitem{Foc28}  M. Born and V. Fock, Z. Phys. \textbf{51}, 165 (1928).
\bibitem{note1}For example, we  consider  a very  simple QCE model consisting
of an ensemble of many identical noninteracting particles in a
one-dimensional power-law potential with size $L$. The power-law
potentials can be parameterized by a single-particle energy spectrum
of the form $E_{n}(L)={E_g(L)} n^{\sigma}$ \cite{W840}, where
$E_g(L)\equiv E_1(L)$ is the energy of the ground state, and
$\sigma$ is the index of the single-particle energy spectrum.
Without loss of generality, the ground state energy  is
   assumed to be proportional to
$L^{-\alpha}$, i.e., $E_g(L)=\gamma L^{-\alpha}$, where $\gamma$ is
a constant for a given potential, and the index $\alpha$ is positive
and depends on the form of the external potential.

\bibitem{note2} For the case with symteric dissipation and without
nonadibatic dissipation, in that work \cite{Esp10} the CA efficiency
was recovered by optimizating power output with respect to time
allocaiton, and thus the correponding time allocation to the hot and
cold reservoirs was obtained.
%\bibitem{Vos85} A. De Vos, Am. J. Phys. \textbf{53}, 570 (1985).
%\bibitem{Bej96} A. Bejan, J. Appl. Phys. \textbf{79}, 1191 (1996).
%\bibitem{Jim07} B. JimenezdeCisneros and A. C. Hernandez, Phys. Rev. Lett. \textbf{98},
%130602 (2007).





\end{thebibliography}
\end{document}